\documentclass[cameraready]{Interspeech}
\usepackage{cite}
\usepackage{amsmath,amssymb,amsfonts}
\usepackage{algorithmic}
\usepackage{graphicx}
\usepackage{textcomp}
\usepackage{multirow}
\usepackage[table,xcdraw]{xcolor}
\usepackage[utf8]{inputenc}

\usepackage{algorithm}

\usepackage{colortbl,xcolor}
\usepackage{tikz}
\usetikzlibrary{patterns}

\newcommand{\patterncellNE}[2][gray!60]{%
  \begin{tikzpicture}[baseline=(X.base)]
    \node[
      inner sep=0pt,
      outer sep=0pt,
      anchor=base,
      text width=32, % requires p{...} columns
      minimum height=2.4ex,
      align=center,
      pattern=north east lines,
      pattern color=#1
    ] (X) {\strut #2};
  \end{tikzpicture}%
}

\definecolor{improvgray}{gray}{0.9} % light grey
\newcommand{\better}[1]{\cellcolor{improvgray}#1} % use for improved results

\definecolor{improvgray}{gray}{0.9}
\def\BibTeX{{\rm B\kern-.05em{\sc i\kern-.025em b}\kern-.08em
    T\kern-.1667em\lower.7ex\hbox{E}\kern-.125emX}}
\title{Group-Aware Partial Model Merging for Children's Automatic Speech Recognition}

\author[affiliation={1}, orcid=0000-0003-2419-1377]{Thomas}{Rolland}
\author[affiliation={1,2}, orcid=0000-0003-2122-5148]{Alberto}{Abad}
\address{
    $^1$ INESC-ID, Portugal \\
    $^2$ Instituto Superior Técnico, Portugal
}

\email{thomas.rolland@inesc-id.pt, alberto.abad@inesc-id.pt}


\usepackage{comment}

\begin{document}

\maketitle

\begin{abstract}
While supervised fine-tuning of adult pre-trained models for children's ASR has shown promise, it often fails to capture group-specific characteristics and variations among children. To address this, we introduce GRoup-Aware PARtial model Merging, a parameter-efficient approach that combines unsupervised clustering, partial fine-tuning, and model merging. Our approach adapts adult-pre-trained models to children by first grouping the children's data based on acoustic similarity. Each group is used to partially fine-tune an adult pre-trained model, and the resulting models are merged at the parameter level. Experiments conducted on the MyST children's speech corpus indicate that GRAPAM achieves a relative WER improvement of 6\%, using the same amount of data, outperforming full fine-tuning while training fewer parameters.
\end{abstract}

\section{Introduction}

Recent advances in Automatic Speech Recognition (ASR) have been made possible by the use of large-scale models, with state-of-the-art results on adult speech corpora \cite{radford2023robust,hsu2021hubert,peng25c_interspeech}. However, performance on children’s speech remains consistently inferior to that on adult speech \cite{sri_end2end, gelin2021endtoend,rolland2024exploring}, primarily due to the pronounced acoustic variability of children’s productions both inter and intra-speaker, arising from ongoing physiological and articulatory development that affects fundamental frequency, formant patterns, and broader spectral–temporal characteristics \cite{Acoustic_change_children,reviewASRchildren}. This difficulty is exacerbated by children’s limited phonetic and linguistic maturity \cite{language_children2}. In addition, progress is further constrained by the scarcity of large, diverse children’s speech corpora, which limits the feasibility of training robust models from scratch \cite{sri_end2end, gelin2021endtoend}. Consequently, existing work has explored augmentation and adaptation strategies such as pitch normalisation \cite{pitchnorm,yadav2019significance}, Vocal Tract Length Normalisation (VTLN) \cite{VTLN2,patel2024improving}, multi-task learning \cite{rolland2022multilingual}, and the use of synthetic speech \cite{wang2021towards,kadyan2021synthesis,rolland2024improved,zhao23c_interspeech}. Among these, supervised fine-tuning (SFT) arise as a widely adopted  strategy \cite{sri_end2end, gelin2021endtoend, fan2022towards, fan2024benchmarking, jain2023adaptation}, adapting an adult-pretrained ASR model to children’s speech via additional training and thereby transferring general acoustic–linguistic representations to children's specifics. Importantly, it was observed that comparable gains can often be obtained by updating only a subset of parameters rather than the full network \cite{rolland2024introduction}, making SFT particularly effective in low-resource child-speech settings where substantial accuracy improvements are achievable with limited data.

Nonetheless, a significant challenge of SFT for children's ASR lies in the substantial acoustic variation across different age groups \cite{1255448,gerosa2009review}. As children’s speech undergoes developmental changes with age, applying a uniform fine-tuning strategy across all age ranges may fail to capture age-specific acoustic characteristics effectively. Recent findings support the use of age-specific ASR models, which consistently yield higher recognition accuracy \cite{lu2022improving,lilles2025fine,hamalainen2014improving}. However, group-specific fine-tuning produce many checkpoints raising concerns surrounding the scalability and parameter efficiency of storing and managing multiple distinct model copies. Additionally, these models typically require prior knowledge of the speaker’s age, which is not always available, making this group-specific training unavailable therefore losing its associated benefits.

Model merging has emerged as a alternative way to combine specialised models trained on different domains without joint retraining, particularly in the domain of large language models (LLMs) \cite{yang2024model,wortsman2022model,yu2024language,jin2022dataless}. 
A key insight from recent work on model merging for LLMs is that the resulting merged model often preserves the capabilities of the individual constituent models \cite{yu2024language,jin2022dataless}. While parameter-level model merging remains a relatively novel area of research, particularly within LLMs and Vision Language Models (VLMs) \cite{chen2025bring,biggs2024diffusion}, emerging studies have begun to explore its potential applicability in ASR \cite{ducorroy25_interspeech,tan2021novo}, including promising developments on children’s speech \cite{shankar2025selective}.

In this work, we explore the potential of model merging as a parameter-efficient strategy to improve ASR for children, with a particular focus on combining models fine-tuned on distinct groups of child speech. The key contributions of this work are as follows:
\begin{enumerate}
    \item We introduce Group-Aware Partial Model Merging (GRAPAM), a novel framework that integrates unsupervised clustering, partial fine-tuning, and model merging to improve children’s ASR.
    \item We conduct a comprehensive analysis of clustering and fine-tuning strategies for effective model merging.
    \item 	We demonstrate consistent improvements over conventional fine-tuning for children's ASR.
    \item We propose and evaluate heterogeneous and iterative merging variants to further enhance performance.
\end{enumerate}

\section{Related work}
\subsection{Model merging}
\label{sec:model_merging}

Model merging has emerged as a promising research direction to combine multiple task-specific models into a single unified architecture that preserve the capabilities of each constituent model \cite{wortsman2022model,matena2022merging,ilharco2022editing}. Unlike traditional multi-task learning, which relies on joint optimisation over multiple datasets, model merging operates directly at the parameter level, enabling the integration of pre-trained models without access to the original training data or additional retraining.

Recent work has proposed diverse model merging strategies. Average Merging, or Linear Interpolation (LERP) averages parameters across fine-tuned models \cite{wortsman2022model}, while Task Arithmetic combines task vectors formed by subtracting a shared base model \cite{ilharco2022editing}. Fisher Merging uses Fisher-information weights to emphasise informative parameters \cite{matena2022merging}, and RegMean casts merging as closed-form linear regression \cite{jin2022dataless}. More recent methods address interference and robustness: TIES trims small-magnitude updates and resolves sign conflicts before merging \cite{yadav2023ties}, and DARE randomly drops and rescales parameters as a pre-merge regulariser \cite{yu2024language}.

Model merging has also been explored for ASR \cite{ducorroy25_interspeech,tan2021novo}. Divide-and-merge (DAM) trains models on data partitions and combines them via genetic search and SGD fine-tuning \cite{tan2021novo}. Recent studies apply merging to Whisper checkpoints for dysarthric speech, improving generalisation in low-resource and long-form settings \cite{ducorroy25_interspeech}, and propose Selective Attention Merging, a layer-wise approach that merges attention layers via task vectors for adult–child transfer, improving children’s ASR \cite{shankar2025selective}.

\subsection{Partial Fine-tuning}
As model scale increases, full SFT becomes increasingly costly and, in low-resource regimes, can degrade performance due to overfitting from updating hundreds of millions of parameters. Parameter-efficient fine-tuning (PEFT) addresses this by adapting models with limited trainable parameters: adapter-based methods \cite{houlsby2019parameter,fan22d_interspeech,10447091} and LoRA \cite{hu2021loralowrankadaptationlarge,10447004} add small trainable modules while keeping the backbone frozen. A closely related strategy, partial fine-tuning (PFT) \cite{ye2023partialfinetuningsuccessorfinetuning,Shen_Liu_Qin_Savvides_Cheng_2021,rolland2024exploring}, instead updates only selected subsets of the original parameters (e.g., attention or feed-forward blocks), leaving the remainder unchanged.
\section{Group-Aware Partial Model Merging}
\begin{figure}
    \centering
    \includegraphics[width=0.6\linewidth]{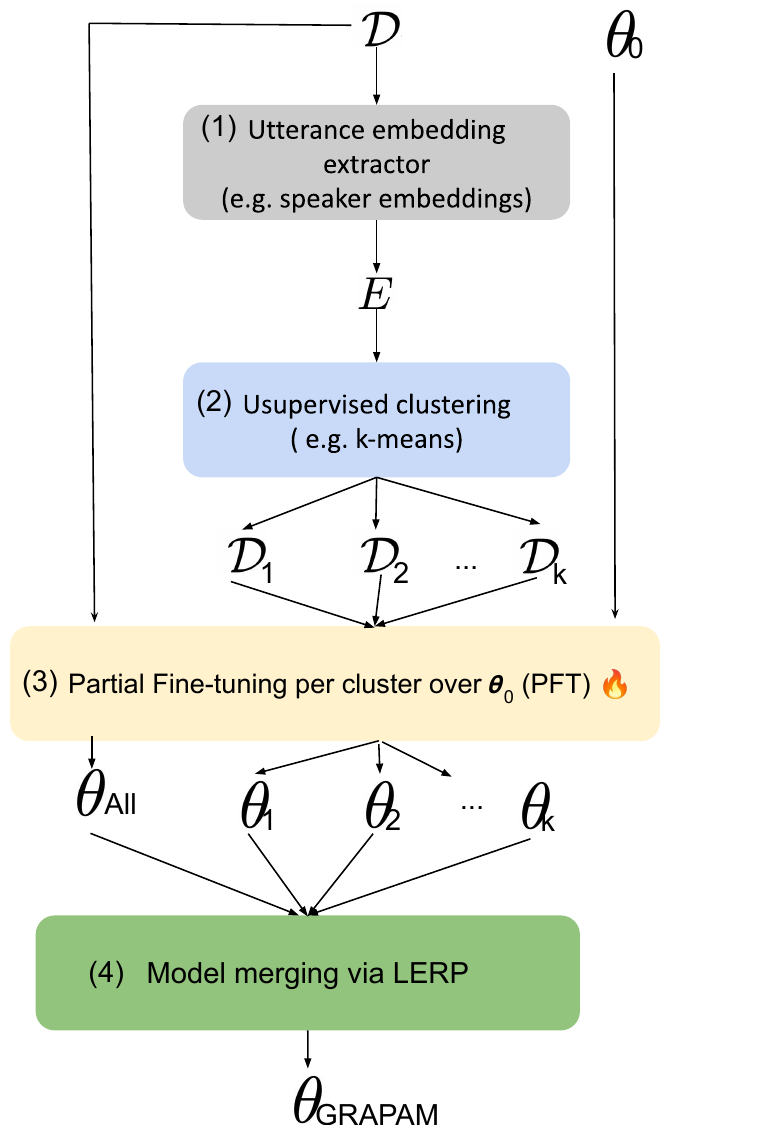}
    \caption{Overview of the four stages of the GRAPAM pipeline. With $\mathcal{D}$ the full dataset and $\theta_0$ the pre-trained adult model parameters. The fire icon denotes stages with ASR fine-tuning.}
    \label{fig:figure}
\end{figure}
We introduce GRAPAM, a novel method that combines insights from several complementary research directions: prior evidence for the effectiveness of age-dependent ASR models \cite{lu2022improving,lilles2025fine,hamalainen2014improving}, recent progress on model merging for children’s ASR \cite{shankar2025selective}, the divide-and-merge paradigm of DAM \cite{tan2021novo}, and advances in partial fine-tuning for parameter-efficient adaptation \cite{rolland2024introduction}. GRAPAM targets children’s speech by mitigating speaker related variability using parameter-efficient fine-tuning followed by model merging. The approach consists of four stages, illustrated in Figure~\ref{fig:figure}.
Let $\mathcal{D}$ denote the training dataset of size $N$, defined as $\mathcal{D} = {(x_i, y_i)}_{i=1}^N$, where $x_i$ represents the input speech signal and $y_i$ denotes the corresponding transcription.

First, $\mathcal{D}$ is partitioned into similarity-based groups. Although age would be a natural criterion, in most children's datasets, age information is not available. Moreover, chronological age may not accurately reflect developmental maturity. To address this, we adopt an unsupervised clustering strategy inspired by prior work \cite{rolland2024exploring}. We extract utterance embeddings $E = \{e_1, e_2, \ldots, e_N\}$, where a single vector represent each utterance. In a second stage, we apply a clustering algorithm to divide these embeddings into K groups. A standard clustering algorithm such as k-means may be employed for this purpose:
\begin{equation}
    \min_{\{C_1, \ldots, C_K\}} \sum_{i=1}^K \sum_{x \in C_i} \lVert x - \mu_i\lVert^2
\end{equation}
where $\mu_i$ is the centroid of cluster $C_i$. Each cluster $C_k$ corresponds to a subset $\mathcal{D}_k \subset \mathcal{D}$ of the training data.

In the third stage, we perform PFT on each group-specific dataset $\mathcal{D}_k$ as well as on the entire dataset $\mathcal{D}$. Given a pre-trained ASR model with parameters $\theta_0$, we selectively fine-tune the feed-forward (FFN) or attention (ATTN) submodules across all layers of the Transformer architecture, resulting in
\begin{equation}
\theta_k = \text{PFT}(\theta_0, \mathcal{D}_k), \quad k = 1, \ldots, K
\end{equation}
and 
\begin{equation}
\theta_{all} = \text{PFT}(\theta_0, \mathcal{D})
\end{equation}

This selective fine-tuning strategy not only ensures parameter efficiency but also facilitates effective adaptation in low-resource scenarios. In this work, we opt for PFT over alternative PEFT approaches such as LoRA or Adapters, as PFT directly modifies existing model parameters. This property significantly simplifies parameter-level merging, which is central to our approach. Consequently, we focus on PFT leaving the exploration of GRAPAM with other PEFT strategies for future work.

In the final stage, we apply LERP to merge the independently fine-tuned models into a single, unified model:
\begin{equation}
\theta_{\text{GRAPAM}} = \alpha_{all} \theta_{all} + \sum_{k=1}^{K}\alpha_k \theta_k
\end{equation}

with constraints that

\begin{equation}  
\alpha_{all} + \sum_{i=1}^{K} \alpha_i = 1, \quad \alpha_{all}\geq 0 \text{ and }  \alpha_i\geq 0 
\end{equation}

In this work, we set the interpolation weights $\alpha_i$ uniformly across all selected fine-tuned models, thereby ensuring that each model contributes equally to the final merged model:
\begin{align}
\alpha_i &= %\frac{1}{|\mathcal{M}|}, & \text{with } i \in \mathcal{M}
\begin{cases}
\frac{1}{|\mathcal{M}|}, & \text{if } i \in \mathcal{M} \\
0, & \text{otherwise}
\end{cases} \notag \\
&\text{for all } i \in \{all,1, \ldots, K\}
\end{align}
with $\mathcal{M} \subseteq \{\theta_{all}, \theta_1,\dots,\theta_k\}$ the set of selected fine-tuned models for merging.
While we use a uniform weight merge strategy, we note that $\alpha_i$ could be further optimised based on validation performance in future work. Finally, the resulting merged model, with parameters denoted as $\theta_{\text{GRAPAM}}$, is used for inference on the entire children’s test set.

\section{Experimental setup}
\subsection{Corpus}
\begin{table}[th]
\caption{My Science Tutor Children Speech Corpus statistics}

\begin{center}
\begin{tabular}{r|c|c|c}
\hline
 & Training & Validation     & Test   \\ \hline
\# of utterances & 42790   & 6812    & 7257  \\ 
\# of speakers & 566   & 80    & 92  \\ 
\# of hours & 117   & 18    & 19  \\ \hline
\end{tabular}
\label{tab:statistics}
\end{center}
\end{table}

For our experiments, we use the My Science Tutor Corpus (MyST) dataset, one of the largest publicly available collections of English children’s speech ($\sim$400 hours). MyST contains child–virtual tutor dialogues spanning eight science domains and includes recordings from 1,372 students in grades 3–5. The dataset is released with predefined partitions that balance scientific domains and ensure that each student appears in only one split. Only 45\% of utterances have word-level transcripts.

We discard utterances shorter than 1 s, dominated by silence, and those longer than 20 s due to GPU constraints. After filtering, the resulting dataset contains 56,859 utterances from 738 speakers, totaling $\sim$154 hours of speech. Detailed partition statistics are reported in Table~\ref{tab:statistics}.

%For our experiments, we used the My Science Tutor Corpus (MyST) dataset. MyST is one of the largest publicly accessible collections of English children’s speech, comprising approximately 400 hours. It encompasses dialogues between children and a virtual tutor across eight scientific domains, involving 1,372 students in grades three to five. The corpus is pre-partitioned, ensuring equitable representation of scientific domains and unique student occurrences within each partition. However, only 45\% of utterances are transcribed at the word level. In this work, we excluded utterances shorter than one second, mainly containing silence, and those longer than 20 seconds, due to GPU constraints. After this filtering process, the dataset comprises 56,859 utterances from 738 speakers, totaling approximately 154 hours of speech. More detailed statistics of the different partitions are provided in Table \ref{tab:statistics}.

\subsection{Implementation details}
All experiments were performed with SpeechBrain \cite{ravanelli2021speechbrain} using Whisper-medium.en \cite{radford2023robust}, a 24-layer encoder–decoder Transformer with 763.9M parameters, pre-trained on $\sim$680k hours of multilingual speech and shown effective for children’s ASR \cite{jain23_interspeech}.

We cluster utterances using three representation types: (i) 192-d speaker embeddings from a pre-trained ECAPA-TDNN model\footnote{https://huggingface.co/speechbrain/spkrec-ecapa-voxceleb}, (ii) 16-d acoustic descriptors obtained by extracting 100 Librosa low-level metrics \cite{mcfee2015librosa} and applying PCA, and (iii) a 1-d WER-based score from Whisper zero-shot predictions. For each representation, we apply k-means from scikit-learn \cite{pedregosa2011scikit} (default K=3); we also report a random clustering baseline.

For PFT, we compare updating (i) all parameters, (ii) attention modules only, or (iii) FFN modules only. Training uses a single RTX A6000 (48GB), batch size 16, learning rate $10^{-5}$, one epoch, and NLL loss. For merging, we use LERP with uniform weights so each cluster contributes equally to the merged model.

\section{Results}
\begin{table*}[th]
\caption{WER (\%) results for GRAPAM. Left: clustering-method variants. Right: partial fine-tuning variants (SFT/PFT). Grey indicates improvement relative to full-model fine-tuning on the entire dataset; diagonal pattern indicates improvement over full SFT but not over the corresponding PFT on $\mathcal{D}$. Best values are in bold.}
\label{tab:all_grapam}
\centering
\setlength{\tabcolsep}{4pt}
\begin{tabular}{|c|c|c|c||cccc||ccc|}
\hline
\multicolumn{4}{|c||}{$\mathcal{M}$} &
\multicolumn{4}{c||}{Clustering method} &
\multicolumn{3}{c|}{Fine-tuning method} \\ \hline
\multicolumn{1}{|c|}{$\theta_{all}$} &
\multicolumn{1}{c|}{$\theta_1$} &
\multicolumn{1}{c|}{$\theta_2$} &
$\theta_3$ &
\multicolumn{1}{c|}{Spk-emb} &
\multicolumn{1}{c|}{Random} &
\multicolumn{1}{c|}{ZS WER} &
\multicolumn{1}{c||}{LSM} &
\multicolumn{1}{c|}{Full SFT} &
\multicolumn{1}{c|}{PFT FFN} &
\multicolumn{1}{c|}{PFT ATTN} \\ \hline

\multicolumn{1}{|c|}{} &
\multicolumn{1}{c|}{} &
\multicolumn{1}{c|}{} &
&
\multicolumn{4}{c||}{14.05} &
\multicolumn{3}{c|}{\phantom{0}14.05} \\ \hline

\multicolumn{1}{|c|}{x} &
\multicolumn{1}{c|}{} &
\multicolumn{1}{c|}{} &
&
\multicolumn{4}{c||}{\phantom{0}9.95} &
9.95 & 9.48 & 9.46 \\ \hline

\multicolumn{1}{|c|}{} &
\multicolumn{1}{c|}{x} &
\multicolumn{1}{c|}{} &
&
10.34 & 10.11 & 10.25 & \better{9.73} &
10.34 & 10.03 & 10.00 \\ \hline

\multicolumn{1}{|c|}{} &
\multicolumn{1}{c|}{} &
\multicolumn{1}{c|}{x} &
&
\better{9.79} & \better{9.69} & 10.32 & 10.61 &
\better{9.79} & \patterncellNE{9.68} & \patterncellNE{9.68} \\ \hline

\multicolumn{1}{|c|}{} &
\multicolumn{1}{c|}{} &
\multicolumn{1}{c|}{} &
x
&
10.36 & 10.63 & 10.18 & 10.03 &
10.36 & \patterncellNE{9.58} & \patterncellNE{9.71} \\ \hline \hline

\multicolumn{1}{|c|}{} &
\multicolumn{1}{c|}{x} &
\multicolumn{1}{c|}{x} &
&
\better{9.48} & 9.91 & 10.52 & 10.12 &
\better{9.48} & \patterncellNE{9.52} & \patterncellNE{9.61} \\ \hline

\multicolumn{1}{|c|}{} &
\multicolumn{1}{c|}{x} &
\multicolumn{1}{c|}{} &
x
&
\better{9.77} & 10.34 & 10.09 & \better{9.45} &
\better{9.77} & \patterncellNE{9.69} & \patterncellNE{9.68} \\ \hline

\multicolumn{1}{|c|}{} &
\multicolumn{1}{c|}{} &
\multicolumn{1}{c|}{x} &
x
&
\better{9.41} & \better{9.74} & \better{9.88} & \better{9.84} &
\better{9.41} & \better{9.44} & \patterncellNE{9.56} \\ \hline

\multicolumn{1}{|c|}{} &
\multicolumn{1}{c|}{x} &
\multicolumn{1}{c|}{x} &
x
&
\better{9.68} & \better{9.68} & 10.01 & \better{\textbf{9.41}} &
\better{9.68} & \better{9.45} & \patterncellNE{9.60} \\ \hline

\multicolumn{1}{|c|}{x} &
\multicolumn{1}{c|}{x} &
\multicolumn{1}{c|}{} &
&
\better{9.72} & 9.98 & 10.32 & \better{9.79} &
\better{9.72} & \patterncellNE{9.66} & \patterncellNE{9.66} \\ \hline

\multicolumn{1}{|c|}{x} &
\multicolumn{1}{c|}{} &
\multicolumn{1}{c|}{x} &
&
\better{9.41} & \better{9.67} & \better{\textbf{9.86}} & \better{9.83} &
\better{9.41} & \better{9.32} & \patterncellNE{9.62} \\ \hline

\multicolumn{1}{|c|}{x} &
\multicolumn{1}{c|}{} &
\multicolumn{1}{c|}{} &
x
&
\better{9.76} & \better{9.77} & 9.96 & \better{9.70} &
\better{9.76} & \patterncellNE{9.66} & \patterncellNE{9.66} \\ \hline

\multicolumn{1}{|c|}{x} &
\multicolumn{1}{c|}{x} &
\multicolumn{1}{c|}{x} &
&
\better{\textbf{9.36}} & \better{\textbf{9.64}} & 9.98 & 9.97 &
\better{\textbf{9.36}} & \better{9.33} & \patterncellNE{9.68} \\ \hline

\multicolumn{1}{|c|}{x} &
\multicolumn{1}{c|}{} &
\multicolumn{1}{c|}{x} &
x
&
\better{9.68} & 9.98 & \better{9.72} & \better{9.68} &
\better{9.68} & \better{9.32} & \patterncellNE{9.66} \\ \hline

\multicolumn{1}{|c|}{PFT} &
\multicolumn{1}{c|}{x} &
\multicolumn{1}{c|}{x} &
x
&
\better{-} & \better{-} & \better{-} & \better{-} &
\better{-} & \better{\textbf{9.36}} & \patterncellNE{{9.51}} \\ \hline

\multicolumn{1}{|c|}{x} &
\multicolumn{1}{c|}{x} &
\multicolumn{1}{c|}{x} &
x
&
\better{9.65} & \better{9.65} & \better{9.93} & \better{9.63} &
\better{9.65} & \better{\textbf{9.31}} & \better{\textbf{9.38}} \\ \hline

\multicolumn{4}{|c||}{Average} &
\better{9.59} & \better{9.84} & 10.03 & \better{9.74} &
\better{9.59} & \better{9.47} & \patterncellNE{9.60} \\ \hline
\end{tabular}
\end{table*}
\subsection{Group-aware model merging}
To assess our approach, we first test whether clustering $\mathcal{D}$ followed by model merging improves children’s ASR, and which utterance representation is most effective. As shown in Table~\ref{tab:all_grapam}, the baseline yields 14.05\% WER, while full-model fine-tuning on $\mathcal{D}$ reduces WER to 9.95\%, confirming that SFT mitigates the adult-speech bias of the pretrained Whisper model.

We then cluster $\mathcal{D}$ using several embedding strategies and merge the resulting domain models according to $\mathcal{M}$. Speaker-embedding (Spk-emb) clustering performs best overall, achieving 9.59\% average WER and a best configuration of 9.36\%. Low-level speech metrics (LSM) is competitive but less consistent (9.74\% average; 9.41\% best), while Random clustering is weaker (9.84\% average; 9.41\% best). Zero-shot WER clustering performs worst (10.03\% average; 9.86\% best), failing to improve over the fine-tuning baseline.

When merging all cluster-specific models ($\mathcal{M}=\{\theta_{all},\theta_1,\theta_2,\theta_3\}$), Spk-emb and Random reach 9.65\% WER and LSM 9.63\%; all remain better than the 9.95\% fine-tuning baseline, supporting model merging as an effective consolidation strategy.

\subsection{Partial model merging}
In Table \ref{tab:all_grapam} we compare three fine-tuning strategies: full-model fine-tuning (Full SFT), partial fine-tuning of FFN blocks (PFT FFN), and partial fine-tuning of attention blocks (PFT ATTN). Fine-tuning on the full dataset $\mathcal{D}$ yields 9.95\% WER with Full SFT, while partial fine-tuning performs better at 9.48\% (FFN) and 9.46\% (ATTN), despite fewer trainable parameters.

Group-aware model merging further reduces WER across configurations, reaching best scores of 9.31\% (PFT FFN) and 9.38\% (PFT ATTN). On average, PFT FFN is strongest (9.47\%), outperforming Full SFT (9.59\%) and PFT ATTN (9.60\%). When merging models trained on the full dataset, fully fine-tuned checkpoints slightly outperform partially fine-tuned ones (9.31\% vs. 9.36\% for PFT–FFN and 9.51\% for PFT–ATTN). The best PFT results arise when merging all cluster models, $\mathcal{M}=\{\theta_{\text{all}},\theta_1,\theta_2,\theta_3\}$.

\subsection{Influence of the number of groups}
\begin{table}[th]
    \centering
\caption{Influence of the number of clusters combination of all fine-tuned model ($\mathcal{M} = \{\theta_{all}, \theta_1,\dots,\theta_k\}$) and the best combination of $\mathcal{M}$ and  results in WER (\%).}
    \label{tab:n_cluster}
    \begin{tabular}{|c|c|c|}
    \hline
    Number of clusters & Combination of all & Best combination \\ \hline
         1 & 9.95 &9.95\\\hline
         2 & \textbf{9.42} &9.41 \\\hline
         3 & 9.65 &9.36 \\\hline
         4 & 9.64 &\textbf{9.33} \\ \hline
    \end{tabular}
    \label{tab:n_cluster}
\end{table}
Table \ref{tab:n_cluster} analyses the effect of the number of clusters (Spk-emb) on GRAPAM, reporting both (i) merging all group models $\mathcal{M}=\{\theta_{all},\theta_1,\dots,\theta_K\}$ and (ii) the best-performing subset of $\mathcal{M}$. The K{=}1 setting (no clustering) matches full-data fine-tuning at 9.95\% WER. For the best subsets, increasing K improves performance, reaching the lowest WER with four clusters (9.33\%). Merging all group models remains competitive, achieving its best result with two clusters (9.42\%). Overall, finer cluster granularity appears to better capture speaker-related variability, yielding gains when the merged model set is appropriately selected.

\subsection{Heterogeneous merging}
Table~\ref{tab:heterogeneous} reports GRAPAM results when merging all cluster models, $\mathcal{M}=\{\theta_{all},\theta_1,\theta_2,\theta_3\}$, across PFT variants and utterance-embedding choices. It also includes heterogeneous settings (ALL row/column), where models trained with different clustering representations and fine-tuning strategies are merged. Overall, PFT–FFN yields the lowest WER (9.36\%). Merging across embedding types remains competitive, with WERs of 9.38\% (speaker embeddings), 9.40\% (LSM), and 9.39\% (random). Importantly, combining models across embeddings or fine-tuning strategies produces only minor changes, indicating that merging largely preserves the underlying capabilities of the constituent models. The full heterogeneous merge (ALL) achieves 9.32\% WER, matching the best individual configurations and demonstrating robustness to heterogeneous merge.

\begin{table}[t]
\caption{WER (\%) results of different combination of PFT and utterance embedding as well as the heterogeneous merging of them. The best-performing combination is shown in bold.}
\label{tab:heterogeneous}
\centering

\begin{tabular}{|c|c|c|c||c|}
\hline
\multicolumn{1}{|l|}{}                 & Full SFT              & PFT FFN               & PFT Attn              & \multicolumn{1}{l|}{ALL} \\
\hline
Spk-emb                   & 9.65                         & \textbf{9.31}                & \textbf{9.38}                & 9.38                      \\
\hline
LSM                             & \textbf{9.63}                         & 9.32                         & 9.45                         & 9.40                      \\
\hline
Random                          & 9.65                         & 9.64                         & \textbf{9.38}                & 9.39                      \\
\hline \hline
ALL & 9.60 & 9.36 & 9.51 & \multicolumn{1}{c|}{9.32}  \\ \hline
                            
\end{tabular}
\end{table}
\vspace{-0.3cm}
\subsection{Iterative Group-Aware Partial Merging }

\begin{table}[t]
\centering
\caption{WER (\%) for different embedding methods (Spk embedding, LSM, Random) across PFT (FFN, ATTN, FULL) and iterations. Selected models as a source for the next turn are presented in bold.}
\label{tab:context_integration_results}
\begin{tabular}{|c|l|c|c|c|}
\hline
PFT & Utt embedding & Iter 1 & Iter 2 & Iter 3 \\ \hline
Partial & - & 9.48 & 10.14 & 10.28 \\ \hline \hline

\multirow{3}{*}{FFN} 
 & Spk embedding & \textbf{9.31} & 9.30 & 9.35 \\ \cline{2-5}
 & LSM           & 9.32 & 9.29 & 9.34 \\ \cline{2-5}
 & Random        & 9.64 & 9.32 & 9.63 \\ \hline

\multirow{3}{*}{ATTN} 
 & Spk embedding & 9.38 & 9.32 & 9.36 \\ \cline{2-5}
 & LSM           & 9.45 & 9.30 & 9.35 \\ \cline{2-5}
 & Random        & 9.38 & 9.33 & 9.35 \\ \hline

\multirow{3}{*}{FULL} 
 & Spk embedding & 9.65 & 9.32 & 9.69 \\ \cline{2-5}
 & LSM           & 9.63 & \textbf{9.28} & \textbf{9.33} \\ \cline{2-5}
 & Random        & 9.65 & 9.88 & 9.92 \\ \hline
\end{tabular}
\end{table}
Table~\ref{tab:context_integration_results} reports WER for successive GRAPAM iterations, where the best configuration from one iteration is used to initialise the next, with $\mathcal{M}=\{\theta_{all},\theta_1,\theta_2,\theta_3\}$. Training the full model for additional epochs on the entire dataset degrades performance, with WER increasing from 9.48\% to 10.28\%, consistent with overfitting.

In contrast, iterative GRAPAM improves recognition: starting from speaker-embedding clustering with FFN partial fine-tuning, the best result is 9.28\%, obtained at iteration 2 after applying LSM-based clustering with full fine-tuning. This gain is consistent with incremental exposure to complementary cluster structures across iterations, which may encourage more robust representations. After iteration 2, performance saturates and no further gains are observed, suggesting limited additional information in the available partitions and potential onset of overfitting. Overall, the results support iterative group-aware model merging as an effective mechanism for improving children’s ASR.

\section{Conclusion and future work}
In this work, we introduced GRAPAM, a parameter-efficient framework for adapting adult-pretrained ASR models to children’s speech via group-aware partial model merging. By integrating unsupervised clustering, partial fine-tuning, and parameter-level interpolation, GRAPAM reduces WER on MyST from 9.95\% to 9.31\%, achieving comparable or improved accuracy without multi-task training or additional data.

Several directions remain for future work. These include replacing uniform interpolation with adaptive weighting, learning merge coefficients on held-out validation data. Exploring stronger merging strategies, including clustering based on self-supervised speech representations and extending GRAPAM to other tasks such as pathological speech.

\section{Acknowledgements}
This work has been submitted to Interspeech 2026. \\
Work supported by Portuguese national funds through Fundação para a Ciência e a Tecnologia (FCT), with references UIDB/50021/2020 and 2022/12328/BD, as well as by the Portuguese Recovery and Resilience Plan (RRP) through project C644865762-00000008 (Accelerat.AI).
\bibliographystyle{IEEEtran}
\bibliography{mybib}

\end{document}